\documentstyle[aps,pre,epsfig,twoside]{revtex}

\begin{document}
\draft

\title{{\bf EVOLUTION OF INDUCED AXIAL MAGNETIZATION IN A TWO-COMPONENT MAGNETIZED PLASMA}}

\author{S. Sarkar,  P. Mukhopadhyay  and  M.  Khan }

\address{{\it Center for plasma studies, Faculty of sciences , Jadavpur University, Calcutta - 700032}}

\author{J.Ortner , M.Steinberg  and  W.Ebeling}

\address{{\it Institut f\"ur Physik, Humboldt Universit\"{a}t zu Berlin,
Invalidenstr. 110, D-10115 Berlin, Germany}}
\date{\today}
\maketitle

\begin{abstract}
In this paper, the  evolution of the induced axial magnetization due to the propagation of an EM- wave
along the static background magnetic field in a two-component plasma has been investigated using the Block
equation. The evolution process induces a strong magnetic anisotropy in the plasma medium, depending
non-linearly on the incident wave amplitude. This induced magnetic anisotropy  can  modify the dispersion
relation of the incident EM-wave, which has been obtained in this paper. In the low frequency Alfven wave
limit, this dispersion relation shows that the resulting phase velocity of the incident wave depends on
the square of the incident wave amplitude and on the static background magnetic field of plasma. The
analytical results are in well agreement with the numerically estimated values in solar corona and
sunspots.
\end{abstract}


\section{Introduction}
The investigation of the propagation of electromagnetic waves is a long studied subject of plasma physics
( see \cite{ABR88} and references therein). In the traditional approach one studies low amplitude waves
propagating in an uncorrelated plasma.  Important information on the properties of plasmas in the linear
response regime can be obtained from the knowledge of the dielectric tensor. The dielectric function of a
magnetized uncorrelated plasma has been extensively studied by Horing \cite{Ho65}. The dispersion of low
amplitude waves or the interaction of low intensity particle beams with plasmas may be studied by
employing the dielectric function. Recently, the stopping power of an uncorrelated plasma has been
investigated
\cite{SO00}. There have been two basic lines beyond the traditional investigations of
electromagnetic modes propagation. One line considers the influence of correlation effects on the plasma
dispersion relations. Recent papers are devoted to the study of the dielectric tensor of correlated
magnetized plasmas and to the investigation of the electromagnetic mode dispersion in coupled magnetized
plasmas \cite{ORT94,TOR98}. The other line is aimed at the investigation of nonlinear effects in
uncorrelated plasmas. A growing number of papers is  dedicated to the study of the propagation of intense
radiation in plasmas (recent works are cited in \cite{BGKS99}).

One of the important area in these investigations is the generation of magnetic fields under the influence
of electromagnetic (EM) waves \cite{StWo72}. One of the sources of the generation of induced magnetization is 
the inverse Faraday effect (IFE).
The induced magnetization from IFE due to propagation of several waves in plasma, has been previously
investigated (see  Ref. \cite{ChKhSa90,ChKhSa91,ChKhSa93,ChKhSa94,ChKhSa96} and references therein). 
This phenomenon arises from magnetic moment per unit volume of the ordered motion of charges of both 
signs, in the presence of an electromagnetic wave propagating in plasma \cite{StWo72,Ch81}. 
This induced field must have axial as well
as lateral component depending on the nature of the wave-wave and wave- particle interactions. For an
elliptically polarized Alfven wave propagating along the static back ground magnetic field in a two
component plasma, this induced magnetization were found to be inversely proportional to the cube of the
ambient magnetic field and the square of the incident wave amplitude, and acts in the direction of the
incident wave propagation. Such effects are expected to be significant in the study of various processes
in the sun and other stars, including pulsars. This effect may be demonstrated in laboratory plasmas.

In this paper, it has been shown that the zero harmonic magnetic moment generated from an elliptically polarized
EM-wave along the direction of its propagation induces strong dc magnetic permeability depending
non-linearly on the incident wave amplitude and acts in the same direction as the induced magnetization.
Moreover, a small perturbation of the self-generated zero harmonic magnetic moment starts to evolve. This
evolution can be investigated by using Block equation model \cite{StVu69}.In this paper this evolution has
been studied and it has been shown that it induces strong magnetic anisotropy in the plane perpendicular
to the direction of the incident wave propagation. This induced magnetic anisotropy  is  evident  from
the  existence  of non-vanishing off-diagonal elements of the magnetic permeability tensor, which also
depend non-linearly on the incident wave amplitude.

In general plasma medium is not a magnetic material. However, the propagation of an incident EM wave in a
plasma generates a self generated uniform magnetization $\vec{M}_a$ which induces a strong magnetic 
permeability in the plasma medium depending nonlinearly on the incident wave amplitude. As magnetic
permeability of a ferromagnet is very large, we can assume our resulting plasma medium as a weakly
ferromagnetic medium in which magnetic permeability is large but not as large as for a ferromagnetic 
material medium. The self generated uniform magnetization may be considered as the ground state 
magnetization. Since we have considered long wave length excitations, a continuum theory is appropriate 
to study the evolution of small perturbation in the ground state magnetization.

Thus a weak ferromagnetic behavior of plasma is expected which can change the orientation of the bulk
magnetization of the plasma that can reduce the mobility of electrons and ions and as a result the
displacement current dominates over the conduction current \cite{StVu69}. This effect modifies the
dispersion characteristics of the incident EM-wave. The dispersion relation of the incident EM-wave in the
resulting plasma medium has been obtained in this paper. In the low frequency Alfven wave limit, it has 
been seen that the phase velocity of the incident Alfven wave in the resulting plasma medium depends on the
static back-ground magnetic  field  of the plasma as well as on the square of the incident wave amplitude. As
the induced magnetization is directly proportional to the square of the incident wave  amplitude,  the
increase  in  the wave amplitude causes  to increase the induced IFE magnetization. This pronounces the
induced magnetic anisotropy, and ultimately inhibits the Alfven wave propagation in the resulting plasma
medium. These results  have been verified numerically both in the Solar corona and Sunspots. 

On the basis of this mechanism many authors have already developed a new mechanism of stabilization of
stimulated Brillouin scattering (SBS) in laser produced plasmas, which is a consequence of the
self-generated magnetic field in the SBS process \cite{ChKhSa94,ChKhSa901}. In that case, a temporally
exponentially growing zero harmonic magnetic field was generated in both axial and lateral directions. The
lateral magnetic field was found to be responsible for the initiation of  magnetic anisotropy in the
plasma medium, which can exponentially reduce the phase velocities of incident and scattered light waves.
However, for an elliptically polarized EM-wave propagating parallel to the static back- ground magnetic
field, a zero harmonic induced axial magnetization is only generated. This axial magnetization in the
ground state cannot induce magnetic anisotropy. The evolution of its linear perturbation induces magnetic
anisotropy in the plasma medium, which has been investigated in this paper by using Block equation
model.\par

In section 2, the dc magnetic permeability induced by the self-generated axial magnetic moment has been
obtained. Its evolution has been studied in section 3. The effect of this evolution on the incident
EM-wave is investigated in section 4. Section 5 describes these results in the low frequency Alfven wave
limit. Calculation of the magnetic moment induced by the Alfven wave is given in the Appendix. Numerical
estimation has been followed by discussion cited in section 6.
\section{DC Magnetic Permeability Induced by Self-Generated Axial Magnetic Moment}

In the classical approximation, the  bulk magnetization present in a magnetic material should be due to
orbital angular momentum of charges, because of the distortion of orbital motion under the inference of EM
fields \cite{ChKhSa90,ChKhSa91,ChKhSa93,ChKhSa94,ChKhSa96,StWo72,Ch81}. When an EM wave propagates along
the static background magnetic field, in a two component plasma, the magnetic moment is generated from the
Inverse Faraday effect (IFE) mechanism along the $z$-direction,
\begin{equation}
\label{2.1} \vec{M}_0 =M_0 \vec{z} \, ,       
\end{equation}                        
which has been presented in the Appendix . This magnetic moment can be expressed in the form
\begin{equation}
\label{2.2} \vec{M}_0 =\left(M_{0x},M_{0y},M_{0z} \right) \, ,        
\end{equation}                                    
where  $M_{0x}=0$, $M_{0y}=0$, and           
\begin{equation}
\label{2.3} M_{0z} =-\frac{n_0 c}{2\omega} \sum_{s=e,i} \, \frac{q_s(\alpha_s+Y_s \beta_s)(\beta_s+Y_s
\alpha_s)}{\left(1-Y_s^2\right)^2}\,,          
\end{equation} 
with
\begin{equation}
\label{2.4} \alpha_s = \frac{q_s a}{m_s \omega c} \,,\, \,\beta_s = \frac{q_s b}{m_s \omega c}  \, ; 
\, \, \, Y_s=\frac{\Omega_s}{\omega}
\, ; \, \, \, \Omega_s=\frac{q_s H_0}{m_s c} \, ,
\end{equation}                 
$q_s, m_s, \Omega_s, (s=e,i)$ are charge, mass and cyclotronfrequencies of electrons and ions,
respectively, $a$ and $b$ are the amplitudes of the incident elliptically polarized EM wave, 
$H_0$ is the static background magnetic field.
 The unperturbed plasma density is given by $n_0=n_{0e}=n_{0i}$ and c is the velocity of
light in vacuum, $\omega$ is the frequency of the incident EM wave, $\vec{z}$ is the direction of incident
wave propagation.

Hence the induced magnetization is,

\begin{equation}
\label{2.5} H_z^{in} = 4\pi M_{0z} = -\frac{4\pi n_0 c}{2\omega} \sum_{s=e,i} \frac{q_s(\alpha_s+Y_s
\beta_s)(\beta_s+Y_s \alpha_s)}{\left(1-Y_s^2\right)^2} \, ,
\end{equation}    
which also acts along the direction of wave propagation. Substituting Eqs.(\ref{2.1},\ref{2.2},\ref{2.3})
in the constitutive relation
\begin{equation}
\label{2.6} \vec{B} = \hat{\mu} \vec{H}\, ,
\end{equation}
with
\begin{equation}
\label{2.7} \vec{B}=\vec{H}+4\pi \vec{M} \, ; \, \, \,\vec{H}=\vec{H}_0 \, ; \, \, \, \vec{M}=\vec{M}_0 \,
; \, \, \, \vec{\mu}=\vec{\mu}_0  \, ,
\end{equation}
we obtain
\begin{equation}
\label{2.8} H_0 +4\pi M_{0z} = \mu_{0z} H_0 \, ,
\end{equation}                                                         
and hence
\begin{equation}
\label{2.9} \mu_{0z}=1-\frac{1}{2} \sum_{s=e,i} \frac{X_s}{Y_s} \frac{(\alpha_s+Y_s \beta_s)(\beta_s+Y_s
\alpha_s)}{\left(1-Y_s^2\right)^2}\, ,
\end{equation}                                      
and $\alpha_s, \beta_s$ are the dimensionless amplitude of the incident EM wave, where
$X_s=\omega_{ps}^2/\omega^2$ and $\omega_{ps}^2=4\pi q_s^2 n_0/m_s$ is the plasma frequency of s-th
species of charges.

This shows that the zero harmonic magnetic moment $\vec{M}_{0}=M_{0z} \vec{z}$ induces a strong dc
magnetic permeability $\mu_{0z}$ depending nonlinearly on the incident wave amplitude and in the
$z$-direction. Thus the resulting plasma medium behaves as a ferromagnetic medium with the IFE
magnetization as the ground state magnetization. In the next section, the dynamics of this self-generated
axial IFE magnetization will be studied.

\section{EVOLUTION OF SELF-GENERATED AXIAL MAGNETIC MOMENT IN A WEAKLY FERRO-MAGNETIC MEDIUM}

From a macroscopic point of view, we may consider the ferromagnetic media as continua characterized
by a magnetic moment density called magnetization. The ground state of a ferromagnet is of uniform 
magnetization at absolute zero temperature. A small disturbance in this magnetization will propagate 
in such a medium and this propagation can be studied by  Block equation model \cite{ChKhSa93},
\begin{equation}
\label{3.1} \frac{d \vec{M}}{dt} =\frac{\gamma}{c} \left(\vec{M} \times \vec{H}_{eff}\right)\, ,
\end{equation}                                                              
where $\vec{M}$ is the bulk magnetization and $\vec{H}_{eff}$ is the effective magnetic field in the
medium , $\gamma$ is the charge to mass ratio and $c$ is the velocity of light in vacuum.

The propagation of an elliptically polarized EM wave  in a two component magnetized plasma induces a zero
harmonic axial magnetic moment from IFE, which generates a nonlinear magnetic permeability in the same
direction. The plasma behaves as a weakly ferromagnetic medium and the induced magnetic moment $\vec{M}_0$
acts as its bulk magnetization. This bulk magnetization is immediately perturbed and the resulting
magnetization  follows the equation
\begin{equation}
\label{3.2} \frac{d \vec{M}_s}{dt} =\frac{\gamma_s}{c} \left(\vec{M} \times \vec{H}_{eff}\right)\, ,
\end{equation}                                                      
where   s=e(electron)/i(ion) and the effective magnetic field $\vec{H}_{eff}$ is the sum of the background
magnetic field $\vec{H}_{0}$ and the $1^{st}$ harmonic magnetic field $\vec{H}_{1}$ of the incident em
wave.
Thus, we have
\begin{equation}
\label{3.3} \vec{H}_{eff}=\vec{H}_{0}+\vec{H}_{1}\, .
\end{equation}                                                    
Moreover, the resulting magnetization,
\begin{equation}
\label{3.4} \vec{M}_{s}=\vec{M}_{0s}+\vec{M}_{1s}\, ,
\end{equation}                                             
where $\vec{M}_{1s}$ is the linearized perturbation of the bulk magnetization $\vec{M}_{0s}$. Both
$\vec{M}_{1s}$ and $\vec{H}_{1}$ satisfy the condition     \begin{equation}
\label{3.5} \mid \vec{M}_{1s} \mid \, \ll \, \mid \vec{M}_{0s} \mid \, ; \, \, \, \mid \vec{H}_{1} \mid \,
\ll \, \mid \vec{H}_{0} \mid \, .
\end{equation}          
$\vec{M}_{0s}$ being the zero harmonic magnetic moment from the orbital motion of sth species of charges
given by
\begin{equation}
\label{3.6} \vec{M}_{0s}=\left(0,0,M_{0s_z} \right)\, ,
\end{equation}                                                            
where               
\begin{equation}
\label{3.7} M_{0s_z}= -\frac{n_0 c}{2\omega} \frac{q_s(\alpha_s+Y_s \beta_s)(\beta_s+Y_s
\alpha_s)}{\left(1-Y_s^2\right)^2}\,
\end{equation}
is independent of both space and time. Here,
\begin{equation}
\label{3.8} \vec{M}_{1s} = \left(M_{1s_x},M_{1s_y},0 \right) \, and \, \, \, \vec{H}_{1} =
\left(H_{1x},H_{1y},0 \right)
\end{equation}                     
are the first order perturbations in $\vec{M}_{s}$ and $\vec{H}_{eff}$, respectively. Using
Eqs.(\ref{3.3},\ref{3.4}) in (\ref{3.2}) and linearizing  we obtain
\begin{equation}
\label{3.9} \frac{d \vec{M}_{1s}}{dt} =\frac{\gamma_s}{c}  \left\{ \left(\vec{M}_{0s} \times
\vec{H}_{1}\right) + \left(\vec{M}_{1s} \times \vec{H}_{0}\right) \right\} \, .
\end{equation}
Substitution of (\ref{3.6}) and (\ref{3.8}) in the RHS of (\ref{3.9}) gives
\begin{equation}
\label{3.10} \dot{M}_{1s_x} = -\omega_{Ms} H_{1y}+\Omega_s M_{1s_y}\, ,
\end{equation}                                            
\begin{equation}
\label{3.11} \dot{M}_{1s_y} = \omega_{Ms} H_{1x}-\Omega_s M_{1s_x}\, ,
\end{equation}
where $\omega_{Ms}=q_s M_{0s_z}/m_s c$ is the magnetization frequency of the sth species of charge
particles, which depends on the induced magnetization $M_{0s_z}$. It is actually the frequency of gyration
of charge particles about the lines of forces of the induced magnetic field $\vec{M}_{0s}$.
Substituting $M_{0s_z}$ from (\ref{3.7}) in $\omega_{Ms}$, we obtain
\begin{equation}
\label{3.12} \omega_{Ms}= -\frac{\omega}{8\pi} \frac{X_s(\alpha_s+Y_s \beta_s)(\beta_s+Y_s
\alpha_s)}{\left(1-Y_s^2\right)^2} \, .
\end{equation}                                    
From (\ref{3.10}) and (\ref{3.11}) we obtain the coupled differential equations,
\begin{equation}
\label{3.13} \left(D^2+\Omega_s^2\right) M_{1s_x} = -\omega_{Ms} \dot{H}_{1y}+\Omega_s \omega_{Ms}
H_{1x}\, ,
\end{equation}                                            
\begin{equation}
\label{3.14} \left(D^2+\Omega_s^2\right) M_{1s_y} = \omega_{Ms} \dot{H}_{1x}-\Omega_s \omega_{Ms} H_{1y}
\,.
\end{equation}
Since $H_{1x}$, $H_{1y}$ are the $x$ and $y$ components of  the magnetic field $\vec{H}_1$ of the incident
EM wave, propagating along the z-direction, and hence $M_{1s_x}$, $M_{1s_y}$ are all proportional to
$\exp[i(kz-\omega t)]$. Hence (\ref{3.13}) and (\ref{3.14}) can be written in the form
\begin{equation}
\label{3.15} M_{1s_x} = \frac{\Omega_s \omega_{Ms}}{\Omega_s^2-\omega^2} H_{1x}+\frac{i \omega
\omega_{Ms}}{\Omega_s^2-\omega^2} H_{1y}\, ,
\end{equation}                         
\begin{equation}
\label{3.16} M_{1s_y} = \frac{-i \omega \omega_{Ms} }{\Omega_s^2-\omega^2} H_{1x}+\frac{ \Omega_s
\omega_{Ms}}{\Omega_s^2-\omega^2} H_{1y}\, ,
\end{equation}
or equivalently,
\begin{equation}
\label{3.17} \vec{M}_{1s} = \hat{\chi}_s \vec{H}_1 \, ,
\end{equation}                                                           
where                           
\begin{equation}
\label{3.18} \hat{\chi}_s =\left( \begin{array}{ccc} \frac{\Omega_s \omega_{Ms}}{\Omega_s^2-\omega^2} &
\frac{i \omega \omega_{Ms} }{\Omega_s^2-\omega^2}&0 \\\frac{-i \omega \omega_{Ms}
}{\Omega_s^2-\omega^2}&\frac{\Omega_s \omega_{Ms}}{\Omega_s^2-\omega^2}&0 \\0&0&0\end{array} \right)\, ,
\end{equation}
is the magnetic susceptibility tensor, whose non vanishing component are
\begin{eqnarray}
\label{3.19}  \chi_{s_{xx}} & = & \chi_{s_{yy}} = \frac{\Omega_s \omega_{Ms}}{\Omega_s^2-\omega^2}\, ,
\nonumber\\
\label{3.20}  \chi_{s_{xy}} & = & \chi_{s_{yx}} = \frac{\omega \omega_{Ms}}{\Omega_s^2-\omega^2}\, .
\end{eqnarray}                                           
Hence the net induced magnetic susceptibility of the resulting plasma medium is
\begin{equation}
\label{3.21}  \hat{\chi} = \hat{\chi}_e+\hat{\chi}_i = \sum_{s=e,i} \hat{\chi}_s\, .
\end{equation}                        
Consequently the induced magnetic permeability of the medium becomes
\begin{equation}
\label{3.22}  \hat{\mu} = \hat{I}+4\pi \sum_{s=e,i} \hat{\chi}_s\, ,
\end{equation}                                                   
where $\hat{I}$ is the unit matrix of order 3. Substitution  of (\ref{3.20}) in (\ref{3.22}) gives
\begin{equation}
\label{3.23}  \mu_{xx} = 1+\sum_{s=e,i} \frac{4\pi \Omega_s \omega_{Ms}}{\Omega_s^2-\omega^2} \, ,
\end{equation}                                 
\begin{equation}
\label{3.24}  \mu_{xy} = -\sum_{s=e,i} \frac{4\pi \omega \omega_{Ms}}{\Omega_s^2-\omega^2} \, ,
\end{equation}                                                      
and hence                                                             
\begin{equation}
\label{3.25}  \mu_{xx} = 1+\frac{1}{2} \sum_{s=e,i} \frac{X_s Y_s (\alpha_s+Y_s \beta_s)(\beta_s+Y_s
\alpha_s)}{\left(1-Y_s^2\right)^3}\, ,
\end{equation}                                     
\begin{equation}
\label{3.26}  \mu_{xy} = -\frac{1}{2} \sum_{s=e,i} \frac{X_s (\alpha_s+Y_s \beta_s)(\beta_s+Y_s
\alpha_s)}{\left(1-Y_s^2\right)^3}\, .
\end{equation}                           
It is seen from the expression (\ref{3.25}) and (\ref{3.26}) that $\mu_{xx}$ and $\mu_{xy}$ depend on the
square of the  incident wave amplitude. The nonvanishing off-diagonal elements $\mu_{xy}$ and $\mu_{yx}$
indicate that a strong magnetic anisotropy is developed in the $xy$-plane, perpendicular to the direction
of incident wave propagation (along the z-direction). This anisotropy is exclusively due to the evolution
of the perturbation $\vec{M}_1$ in a plasma medium, having weakly ferromagnetic properties.

\section{EFFECT OF MAGNETIC ANISOTROPY ON THE INCIDENT EM WAVE PROPAGATION}
As we are studying magnetic moment dynamics in the plasma medium under the action of long range coulomb
forces between the charge particles, classical theory is more appropriate than quantum mechanical theory
because any disturbance propagates through a plasma medium with a wave length much greater than the atomic
distances.

In this section we shall investigate how this induced magnetic anisotropy changes the dispersion
characteristics of the incident elliptically polarized EM wave.

The induced magnetic anisotropy reduces the mobility of the charge particles. Hence the conduction current
becomes negligible and the displacement current dominates over conduction current. 
Thus in such an insulated ferromagnet, the
propagation of an EM wave obeys the following Maxwell equations
\begin{equation}
\label{4.1}  \vec{\nabla} \times \vec{E} = -\frac{1}{c} \frac{\partial \vec{B}}{\partial t}\, ,
\end{equation}                                                       
\begin{equation}
\label{4.2}  \vec{\nabla} \times \vec{H} = \frac{1}{c} \frac{\partial \vec{D}}{\partial t}\, ,
\end{equation}                                                        
with constitutive relations
\begin{eqnarray}
\label{4.3}  \vec{D} & = & \hat{\epsilon} \vec{E}\, , \nonumber\\
\label{4.4}  \vec{B} & = & \hat{\mu} \vec{H}\, ,
\end{eqnarray}                                                              
where $\hat{\epsilon}$, $ \hat{\mu}$ are respectively the dielectric tensor and magnetic permeability
tensor of the resulting plasma medium, $\hat{\mu}$ has been already obtained in section 3. Since the
plasma under  consideration is initially magnetized, it has a dielectric anisotropy of the form
\begin{equation}
\label{4.5} \hat{\epsilon} =\left( \begin{array}{ccc} \epsilon_{xx}&-i\epsilon_{xy}&0 \\
i\epsilon_{xy}&\epsilon_{yy}&0 \\0&0&0\end{array} \right)\, ,
\end{equation}
where  
\begin{eqnarray}
\label{4.6}  \epsilon_{xx}=1-\sum_{s=e,i} \frac{X_s}{1-Y_s^2}\, , \nonumber\\
\label{4.7}  \epsilon_{xy}=\sum_{s=e,i} \frac{X_s Y_s}{1-Y_s^2}\, .
\end{eqnarray}
Since the electric field $E_{1x}$, $E_{1y}$ and magnetic field $H_{1x}$, $H_{1y}$ of the incident
electromagnetic wave contain the phase factor $\exp[i(kz-\omega t)]$, where $\omega$ and $k$ are its
frequency and wave number, the Maxwell equations (\ref{4.1}) and (\ref{4.2}) together with the state
relations (\ref{4.4}) reduce to
\begin{equation}
\label{4.8} n \, \left( \begin{array}{c} E_{1x} \\ E_{1y} \end{array}\right)  =\left( \begin{array}{cc} i
\mu_{xy}&\mu_{yy} \\ -\mu_{xx}&i\mu_{xy} \end{array} \right) \, \left( \begin{array}{c} H_{1x} \\ H_{1y}
\end{array}\right)\, ,
\end{equation}
\begin{equation}
\label{4.9} n \, \left( \begin{array}{c} H_{1x} \\ H_{1y} \end{array}\right)  =\left( \begin{array}{cc} i
\epsilon_{xy}&\epsilon_{yy} \\ -\epsilon_{xx}&i\epsilon_{xy} \end{array} \right) \, \left(
\begin{array}{c} E_{1x} \\ E_{1y} \end{array}\right)\, .
\end{equation}                                      
Substituting $E_{1x}$, $E_{1y}$, $H_{1x}$, $H_{1y}$ from (\ref{A.15}) and(\ref{A.16}), we obtain
\begin{equation}
\label{4.10}  n^2 = \left(  \epsilon_{xx} \pm \epsilon_{xy}\right)\, \left( \mu_{xx} \pm \mu_{xy}\right)
\, ,
\end{equation}                                               
which is  the modified dispersion relation of the incident EM wave in the anisotropic plasma medium.
Substitution of $\mu_{xx}$, $\mu_{xy}$, $\epsilon_{xx}$, $\epsilon_{xy}$  from (\ref{3.25},\ref{3.26}) and
(\ref{4.7}) in (\ref{4.10}) gives
\begin{equation}
\label{4.11}  n^2 = \left(1-\sum_{s=e,i} \frac{X_s}{1 \pm Y_s}\right)  \left(1 \mp \frac{1}{2}
\sum_{s=e,i} \frac{X_s (\alpha_s+Y_s \beta_s)(\beta_s+Y_s \alpha_s)}{(1 \pm Y_s)
\left(1-Y_s^2\right)^2}\right)\, ,
\end{equation} 
where $n=kc/\omega$ is the
refractive index of the resulting plasma medium. This dispersion relation shows that the dispersion
characteristics of the incident EM wave depend on the product of the incident EM wave amplitude. In the
next section we shall investigate such characteristics for the case of Alfven waves.

\section{ALFVEN WAVE APPROXIMATION}
If the incident EM wave propagating in a two component plasma along the static background magnetic field
is an Alfven wave, the wave frequency satisfies the condition
\begin{equation}
\label{5.1} \omega \ll \Omega_{e}, \Omega_i \, .              
\end{equation}                                                
                                      
Under the approximation (\ref{5.1}) the dispersion relation (\ref{4.11}) reduces to
\begin{equation}
\label{5.3} n^2_\pm=\left(\frac{k^2 c^2}{ \omega^2}\right)_{\pm} =\left(1+\frac{c^2}{c_A^2}\right)\,
\left(1- \frac{c^2}{c_A^2} \frac{a^2\mp ab+b^2}{H_0^2} \right) \, ,              
\end{equation}
where $c_A^2= H_0^2/4\pi n_0 m_i$ is the Alfven velocity in the plasma. Thus we get two branches of mode
propagation. In the case of a low amplitude wave, $a,b \to 0$, we obtain from Eq. (\ref{5.3}) the
dispersion relation of the ordinary Alfven wave with only one branch. Consider the modifications of  wave
propagation caused by the nonlinearity. First, the nonlinearity produces a splitting of the Alfven branch
into two branches  corresponding to the left elliptically polarized ($a$ and $b$ of equal sign) or to the
right elliptically polarized Alfven wave ($a$ and $b$ of different sign), respectively. Second, the phase
velocity of the incident Alfven wave in the resulting anisotropic plasma medium becomes
\begin{equation}
\label{5.4} \left(\frac{\omega}{k}\right)_{\pm} = \pm \frac{c_a c}{\sqrt{c_A^2+c^2}}/ \sqrt{1-
\frac{c^2}{c_A^2} \frac{a^2\mp ab+b^2}{H_0^2} }\, .              \end{equation}                                                  
This shows that Alfven wave can propagate with very large but finite phase velocity (\ref{5.4}), if
\begin{equation}
\label{5.5} \frac{c_A^2}{c^2}\frac{H_0^2}{a^2\mp ab+b^2} > 1 \, .             
\end{equation}
Only very long wave length Alfven waves can propagate, if
\begin{equation}
\label{5.6} \frac{c_A^2}{c^2}\frac{H_0^2}{a^2\mp ab+b^2} \cong 1 \, ,           \end{equation}                                                        
The first brunch of the nonlinear Alfven wave has a cutoff, if        
\begin{equation}
\label{5.7} \frac{c_A^2}{c^2}\frac{H_0^2}{a^2+ ab+b^2} <1 \, ,          \end{equation}
the second brunch has ist cutoff, if
\begin{equation}
\label{5.8} \frac{c_A^2}{c^2}\frac{H_0^2}{a^2- ab+b^2} <1 \, .          \end{equation}
We see that for a given amplitude of the Alfven wave the magnetic field strength $H_0$ should exceed a
certain threshold $H_0 > \left(4 \pi n_0 m_i c^2 (a^2 \pm ab + b^2) \right)^{1/4}$ to make propagation of
the Alfven wave possible.
                                                                
\section{Numerical Estimation}
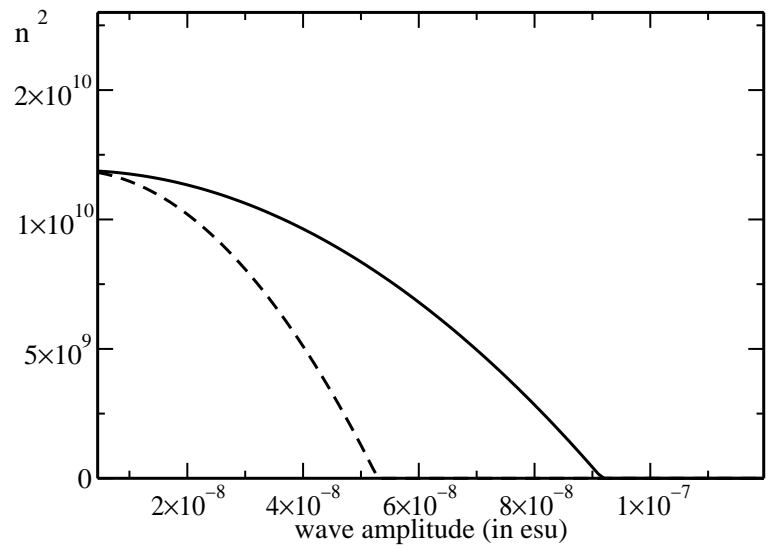
\begin{figure}[h]
\centerline{\epsfig {figure=alfven.eps,width=10.0cm,angle=0}}
\caption{}\label{alfven}
\end{figure}

In the solar corona, the ambient magnetic field is, $H_0=10^{-2} {\rm G}$  and the plasma mass density is,
$\rho = 10^{-16} {\rm g}/{\rm cm}^3$. Hence the Alfven speed in  a solar corona is $c_A=2.75 \times 10^5
{\rm cm}/{\rm sec}$. If an Alfven wave of amplitude $a=10^{-8} {\rm esu}$ propagates in the solar corona,
the induced magnetic field would be of the order of $10^{-4} {\rm G}$ and the phase velocity of the
incident wave in the resulting anisotropic plasma medium is very large but finite. This implies that only
very long wave length waves can propagate in such a medium. For an  incident wave amplitude $a=10^{-7}
{\rm esu}$, the induced magnetization is $10^{-2} {\rm G}$ and the wave phase velocity $\omega/k$ becomes
infinite. Moreover, if the incident wave amplitude increases to a value $10^{-6} {\rm esu}$, the induced
magnetization becomes $1.13 {\rm G}$ and the phase velocity of the incident wave becomes imaginary and no
further wave propagation is possible. The refractive index of nonlinear circular polarized Alfven wave
with amplitude $a=b$ and propagating in the solar corona is shown in Fig. \ref{alfven}. With increasing
wave amplitude the refractive indices of  both nonlinear Alfven brunches decrease. The first brunch
($n_+^2$) has its cutoff at an amplitude $a=5.3 \cdot 10^{-8} {\rm esu}$, whereas the second brunch ranges
up to an amplitude of $a=9.1 \cdot 10^{-8} {\rm esu}$.

Similar results have also been obtained in sunspots, where $H_0 = 3000 {\rm G}$, mass density $\rho
=10^{-5} {\rm g}/{\rm cm}^3$ and Alfven speed $c_A = 2.7 \times 10^5 {\rm cm}/{\rm sec}$. In that case for
$a=10^{-2} {\rm esu}$, $H^{in} = 420 {\rm G}$ and $\omega/k$ is very large but finite. For $a=10^{-1.5}
{\rm esu}$, $H^{in}$  becomes $4200 {\rm G}$ and $\omega/k$ is infinite, and for $a=10^{-1} {\rm esu}$,
$H^{in} = 4.2 \times 10^4 {\rm G}$, $\omega/k$ is imaginary.

These numerical results confirm that the increase in the incident wave amplitude produces a magnetic
anisotropy via the increasing induced magnetization in the plasma medium and consequently inhibits the
Alfven wave propagation.

\section{DISCUSSION}
From the results so far obtained in this paper, it is evident that for a strong EM wave propagating along
the static back ground magnetic field in a two component plasma, the self generated zero harmonic axial
magnetic moment starts to evolve. This evolution induces a strong magnetic anisotropy in the plasma medium
and the medium consequently behaves as a weakly ferromagnetic medium with the zero harmonic magnetization
as the ground state magnetization. This anisotropy inhibits the incident wave propagation in the resulting
plasma medium. Moreover, as the wave amplitude increases, the anisotropy becomes strong and absorption of
the wave by the medium is pronounced.

\begin{appendix}
\section{CALCULATION OF THE MAGNETIC MOMENT FOR AN  ELLIPITICALLY POLARIZED EM-WAVE IN A TWO-COMPONENT
MAGNETIZED PLASMA}

We consider the propagation of a transverse EM-wave in a two component cold magnetized plasma  in which
electrons and ions are both
mobile. Collisions have been neglected. The basic equations describing such a plasma model, in the cold
plasma limit are
\begin{equation}
\label{A.1} \frac{\partial \vec{u}_s}{\partial t} +\left(\vec{u}_s \cdot \vec{\nabla} \right) \vec{u}_s =
\frac{q_s}{m_s} \vec{E} + \frac{q_s}{m_s c} \left(\vec{u}_s \times \vec{H} \right) \, ,
\end{equation}                        
\begin{equation}
\label{A.2} \frac{\partial n_s}{\partial t} + \vec{\nabla} \cdot \left(n_s \vec{u}_s \right)    =  0 \, ,     
\end{equation}
\begin{equation}
\label{A.3}  \vec{\nabla} \times \vec{E} = -\frac{1}{c} \frac{\partial \vec{H}}{\partial t}\, ,
\end{equation}                                                       
\begin{equation}
\label{A.4}  \vec{\nabla} \times \vec{H} = \frac{1}{c} \frac{\partial \vec{E}}{\partial t} +\frac{4\pi}{c}
\vec{j}\, ,
\end{equation}                                                        
\begin{equation}
\label{A.5}  \vec{\nabla} \cdot \vec{E} = 4\pi \rho \, ,
\end{equation}
\begin{equation}
\label{A.6}  \vec{\nabla} \cdot \vec{H}  =  0 \, ,
\end{equation}                                
where $\rho = \sum_{s=e,i} n_s q_s$ and $\vec{j} = \sum_{s=e,i} n_s q_s \vec{u}_s$ are the charge and
current densities in the plasma. Assuming the plasma is initially quasi-static and quasi-neutral, such
that
\begin{equation}
\label{A.7}  \vec{u}_{s0} =0 \, ; \, \, \, n_0=n_{0e}=n_{0i}\, .
\end{equation}                                                
We linearize the field variables,
\begin{equation}
\label{A.8}  \vec{u}_{s} = \vec{u}_{s0}+ \vec{u}_{s1}\, ; \, \, \, n_s=n_{s0}+n_{s1}\, ; \, \, \, \vec{E}
= \vec{E}_{0}+ \vec{E}_{1} \, ; \, \, \, \vec{H} = \vec{H}_{0}+ \vec{H}_{1} \, ,
\end{equation}                         
where $\vec{H_0} = (0,0,H_0)$ is the ambient magnetic field  acting along the $z$-direction and
$\vec{u}_{s1}$, $n_{s1}$, $\vec{E}_{1}$, $\vec{H}_{1}$ are the first order perturbations in the field
variables about their equilibrium value.

Linearizing the basic equations (\ref{A.1},\ref{A.2},\ref{A.3},\ref{A.4},\ref{A.5},\ref{A.6}) with the
help of (\ref{A.8}), we obtain
\begin{equation}
\label{A.9} \frac{\partial \vec{u}_{s1}}{\partial t} = \frac{q_s}{m_s} \vec{E}_1 + \frac{q_s}{m_s c}
\left(\vec{u}_{s1} \times \vec{H}_0 \right) \, ,
\end{equation}                        
\begin{equation}
\label{A.10} \frac{\partial n_{s1}}{\partial t} + \vec{\nabla} \cdot \left(n_0 \vec{u}_{s1} \right)     =
0 \, ,     
\end{equation}
\begin{equation}
\label{A.11}  \vec{\nabla} \times \vec{E}_1 = -\frac{1}{c} \frac{\partial \vec{H}_1}{\partial t}\, ,
\end{equation}                                                       
\begin{equation}
\label{A.12}  \vec{\nabla} \times \vec{H}_1 = \frac{1}{c} \frac{\partial \vec{E}_1}{\partial t}
+\frac{4\pi n_0}{c} \sum_{s=e,i} q_s \vec{u}_{s1}\, ,
\end{equation}                                                        
\begin{equation}
\label{A.13}  \vec{\nabla} \cdot \vec{E}_1 = 4\pi \sum_{s=e,i} q_s n_{s1} \, ,
\end{equation}
\begin{equation}
\label{A.14}  \vec{\nabla} \cdot \vec{H}_1  =  0 \, .
\end{equation}                                
Let the first order electric field of the em wave that induces the perturbation in the plasma model be
\begin{equation}
\label{A.15}  \vec{E}_1 = (a \cos \theta, b \sin \theta, 0) \, ; \, \, \, \theta= kz-\omega t\, ,
\end{equation}                                
where $\vec{z}$ being the direction of propagation of the em wave. Substitution of (\ref{A.5}) in
(\ref{A.11}) gives
\begin{equation}
\label{A.16}  \vec{H}_1 = n (-b \sin \theta, a \cos \theta, 0) \, ,
\end{equation}                                        
where $n=kc/\omega$ is the refractive index of plasma, $\omega$  and $k$ are the frequency and wave number
of the incident em wave.
Solution of the linearized set of equations
(\ref{A.9},\ref{A.10},\ref{A.11},\ref{A.12},\ref{A.13},\ref{A.14}) with the help of (\ref{A.15}) and
(\ref{A.16}) gives the first order perturbed velocity of the charged particles $\vec{u}_{s1}$, induced by
the first order transverse em wave
\begin{equation}
\label{A.17}  \vec{u}_{s1} = (u_{s1_x},u_{s1_y}, 0) \, ,
\end{equation}
\begin{equation}
\label{A.18}  \vec{u}_{s1_x} = - \frac{q_s}{m_s} \frac{\omega a +\Omega_s b}{\omega^2-\Omega_s^2} \sin
\theta\, ,
\end{equation}
\begin{equation}
\label{A.19}  \vec{u}_{s1_y} = \frac{q_s}{m_s} \frac{\omega b +\Omega_s a}{\omega^2-\Omega_s^2} \cos
\theta\, .
\end{equation}                                                
After integrating (\ref{A.18}) and (\ref{A.19}), we obtain the first order wave induced displacement of
the charge particles
\begin{equation}
\label{A.20}  \gamma_{s1_x} = - \frac{q_s}{m_s \omega} \frac{\omega a +\Omega_s b}{\omega^2-\Omega_s^2}
\cos \theta\, ,
\end{equation}
\begin{equation}
\label{A.21}  \gamma_{s1_y} = \frac{q_s}{m_s \omega} \frac{\omega b +\Omega_s a}{\omega^2-\Omega_s^2} \sin
\theta\, .
\end{equation}                                                      
Hence the magnetic moment induced by the electrons and ion motion under the influence of the incident EM
wave is
\begin{equation}
\label{A.22}  \vec{M} = \sum_{s=e,i} \vec{M}_s\, ,
\end{equation}                                        
where
\begin{equation}
\label{A.22a}  \vec{M}_s = \frac{1}{2c} \left( \vec{\gamma}_{s1} \times \vec{j}_{s1}  \right)\, ,
\end{equation} 
with $\vec{j}_{s1}=n_0 q_s \vec{u}_{s1}$, is the first order perturbed current density due to the wave
induced motion of charge particles.
From (\ref{A.18},\ref{A.19},\ref{A.20},\ref{A.21},\ref{A.22}) we obtain
\begin{equation}
\label{A.23}  \vec{M} = \left( 0,0,M_z\right)\, ,
\end{equation}                                                        
where                         
\begin{equation}
\label{A.24}  M_z =-\frac{n_0 c}{2\omega} \sum_{s=e,i} \frac{q_s (\alpha_s+Y_s \beta_s)(\beta_s+Y_s
\alpha_s)}{\left(1-Y_s^2\right)^2}\, ,
\end{equation}
where $(\alpha_s,\beta_s)=q_s(a,b)/m_s \omega c$, $Y_s=\Omega_s/\omega$, $\Omega_s=q_sH_0/m_s c$. Here
$q_s$ and $m_s$ are charge and mass of the sth species of charge particles.

This is the induced magnetization from IFE generated from the distortion of the ordered motion of charge
particles under the influence of incident EM wave. Eq. (\ref{A.23}) shows that this induced magnetic
moment acts along the $z$-direction which is the direction of the incident wave propagation. This is the
ground state magnetization $\vec{M}=\vec{M_0}$ of the weakly ferromagnetic medium as discussed in this
paper.
Hence the induced magnetization $\vec{H}^{in}$ is
\begin{equation}
\label{A.25}  \vec{H}^{in} =4\pi \vec{M}   \, .
\end{equation}                                                        
Substituting (\ref{A.23}) and (\ref{A.24}) in (\ref{A.25}), we obtain
\begin{equation}
\label{A.26}  \vec{H}^{in} =\left(0,0,H_z^{in} \right)  \, ,
\end{equation}                                                
where,
\begin{equation}
\label{A.27}  H_z^{in} =  -\frac{2\pi n_0 c}{\omega} \sum_{s=e,i} \frac{q_s (\alpha_s+Y_s
\beta_s)(\beta_s+Y_s \alpha_s)}{\left(1-Y_s^2\right)^2}\, .
\end{equation}

\end{appendix}        

\newpage

Figure Captions\\

Fig. 1 Squares of refraction indices $n_\pm^2$ of nonlinear left circular polarized Alfven wave
propagating in the solar corona vs. wave amplitude $a$ (in esu): solid line - $n_-^2$, dashed line -
$n_+^2$. Magnetic field strength is $H_0=10^{-2} {\rm G}$  and plasma mass density is $\rho = 10^{-16}
{\rm g}/{\rm cm}^3$.

\newpage

\begin{thebibliography}{99}

\bibitem{ABR88}  Aleksandrov, A.F., Bogdankevich, L.S. and Rukhadze, A.A.
\textit{ Fundamentals of Plasma Electrodynamics, } Vyssh. Shkola, Moscow (1988) [in
Russian].

\bibitem{Ho65}
N.J.Horing, Ann. Phys. (N.Y.) {\bf 31}, 1 (1965).

\bibitem{SO00}
 M. Steinberg and J. Ortner, Phys. Rev. E {\bf 63}, 046401 (2001).



\bibitem{ORT94}
J.Ortner, V.M.Rylyuk, and I.M. Tkachenko,  Phys. Rev. E, {\bf 50}, 4937 (1994).

\bibitem{TOR98}
I.M. Tkachenko, J.Ortner, and V.M.Rylyuk,  Phys. Rev. E, {\bf 57}, 4846 (1998).

\bibitem{BGKS99} A.V.~Borovsky, A.~L.~Galkin, V.~V.~Korobkin and O.~B.~Shiryaev, Phys. Rev. E {\bf 59},
2253 (1999).

\bibitem{StWo72}
A.D. Steiger and C.H. Woods, Phys. Rev. A {\bf 5}, 1467 (1972).


\bibitem{ChKhSa90}
B. Chakraborty, M. Khan, S. Sarkar, V. Krishan, B. Bhattacharyya, Ann. Phys. (N.Y.) {\bf 201}, 1 (1990).

\bibitem{ChKhSa91}
S. Sarkar, B. Bera, M. Khan, B. Chakraborty, Aust. J. Phys. {\bf 44}, 59 (1991).

\bibitem{ChKhSa93}
B. Chakraborty, S. Sarkar, C. Das, B. Bera, M. Khan, Phys. Rev. E {\bf 47}, 2736 (1993).

\bibitem{ChKhSa94}
S. Sarkar, B. Chakraborty, M. Khan, Phys. Rev. E {\bf 50} , 1458 (1994).

\bibitem{ChKhSa96}
S. Sarkar, D. Dutt, B. Chakraborty, M. Khan, Il Nuovo cimento {\bf 18D}, 75 (1996).


\bibitem{Ch81}
A.C.L. ChaiSn, Phys. Fluids {\bf 24}, 369 (1981).

\bibitem{StVu69}
M.C. Steele and B. Vural, Wave Interaction in Solid State Plasma, McGraw Hill P.167 (1969).

\bibitem{ChKhSa901}
M. Khan, S. Sarkar, T. Desai, H.C. Pant, Laser and Particle Beams. 16 , 491 (1998).





\end{thebibliography}
\end{document}